\documentclass[twocolumn,showpacs,preprintnumbers,amsmath,amssymb]{revtex4}

\usepackage{graphicx}
\usepackage{dcolumn}
\usepackage{bm}
\usepackage[usenames]{color}

\begin{document}

\preprint{APS/123-QED}

\title{Electronic transport properties of a tilted graphene pn junction  }

\author{Tony Low and Joerg Appenzeller}

\affiliation{$^{1}$School of Electrical \& Computer Engineering, Purdue University, West Lafayette, IN47906, USA }%

\date{\today}
\email{tonyaslow@gmail.com}
\begin{abstract}
Spatial manipulation of current flow in graphene could be achieved through the use of a tilted pn junction. We show through numerical simulation that a pseudo-Hall effect (i.e. non-equilibrium charge and current density accumulating along one of the sides of a graphene ribbon) can be observed under these conditions. The tilt angle and the pn transition length are two key parameters in tuning the strength of this effect. This phenomenon can be explained using classical trajectory via ray analysis, and is therefore relatively robust against disorder. Lastly, we propose and simulate a three terminal device that allows direct experimental access to the proposed effect. 
\end{abstract}

\maketitle
\section{\label{sec:level1}INTRODUCTION}

\begin{figure}[t]
\centering
\scalebox{0.45}[0.45]{\includegraphics*[viewport=120 155 700 500]{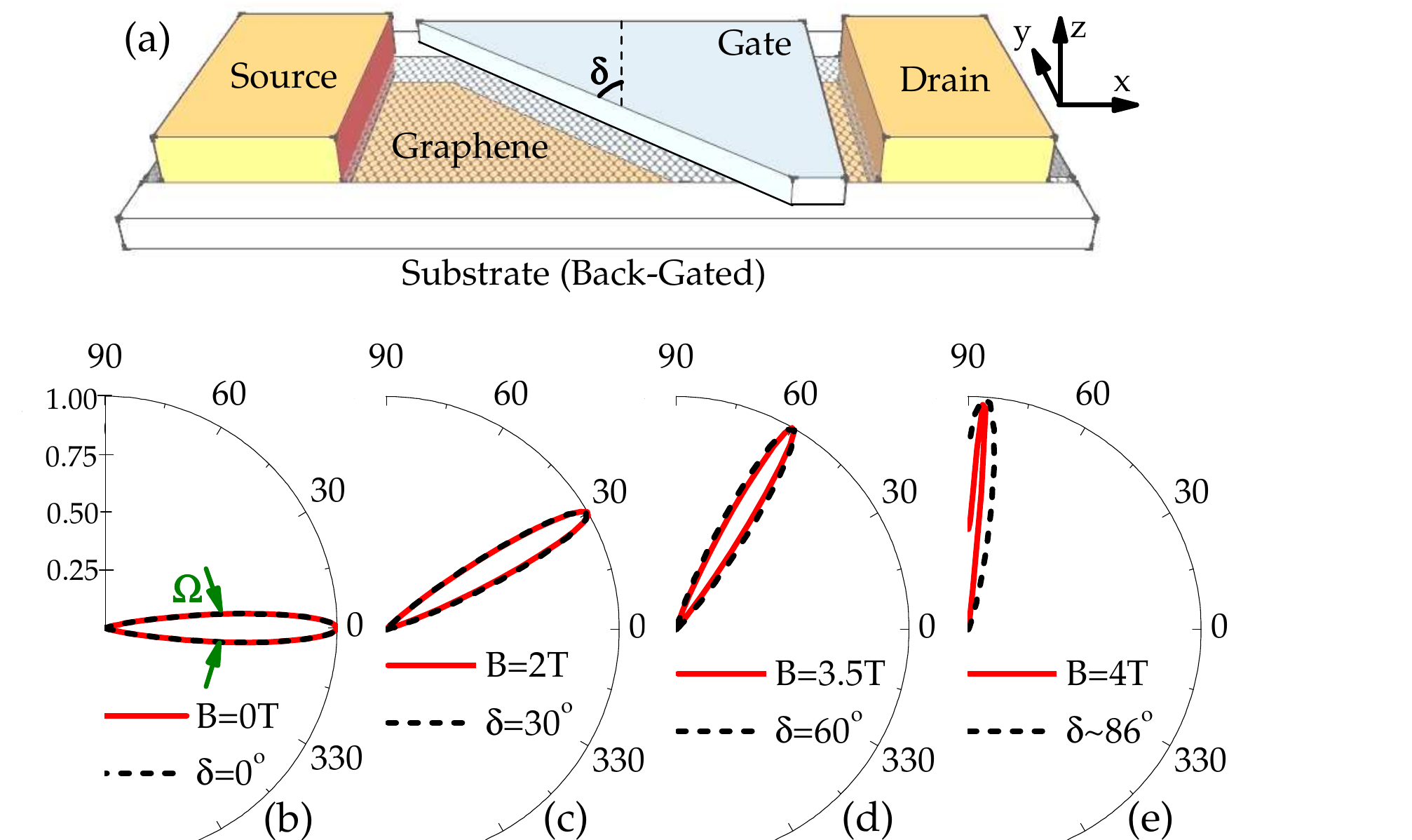}}
\caption{\footnotesize (a) Schematic of a tilted pn junction device built on a graphene ribbon. The top and bottom gate allows the tunability of the electron/hole carrier density on each side of the junction. Tilt angle defined as $\delta$. (b-e) Polar plots of the carrier transmission probability across a symmetric graphene pn junction for different values of magnetic field using the WKB model outlined in \cite{shytov07} (solid lines). The calculation is done for an experimentally typical pn junction with transition width of 100nm and a built-in potential of $0.4eV$ i.e. $\epsilon_{f}=\pm 0.2eV$ for the n/p regions \cite{stander09}. Similar plots for different $\delta$ at $B=0T$ using WKB are also shown (dashed lines).}
\label{figure1}
\end{figure}

A semiconductor pn junction where both sides of the junction are biased such that their Fermi surfaces are identical could potentially serve as an electron analogue of the Veselago lens \cite{veselago68}. Graphene, a zero band-gap two-dimensional semiconductor with Dirac-type linear energy dispersion \cite{novoselov04,novoselov05,zhang05,semenoff84}, is an ideal medium for realizing this physical analogy. As recently advocated by Cheianov et al. \cite{cheianov07c}, an abrupt and symmetrically biased graphene pn junction could function as an electron focusing device. An electronic superlattice of cascading pn junctions could serve as an electron beam colliminator as elaborated by Park et al. \cite{park08}.  In this letter, we propose utilizing a tilted pn junction to manipulate the current flow such that charge carriers preferably propagate along one edge of the sample using a set-up as illustrated in Fig. \ref{figure1}a. This is achieved by controlling the following interface properties: (i) tilt angle $\delta$ and (ii) the extent of the pn transition region, $D$. The possibility to manipulate the spatial distribution of the current density in graphene opens the door for novel electronic device concepts.

\begin{figure*}[htps]
\centering
\scalebox{0.69}[0.69]{\includegraphics*[viewport=38 393 748 560]{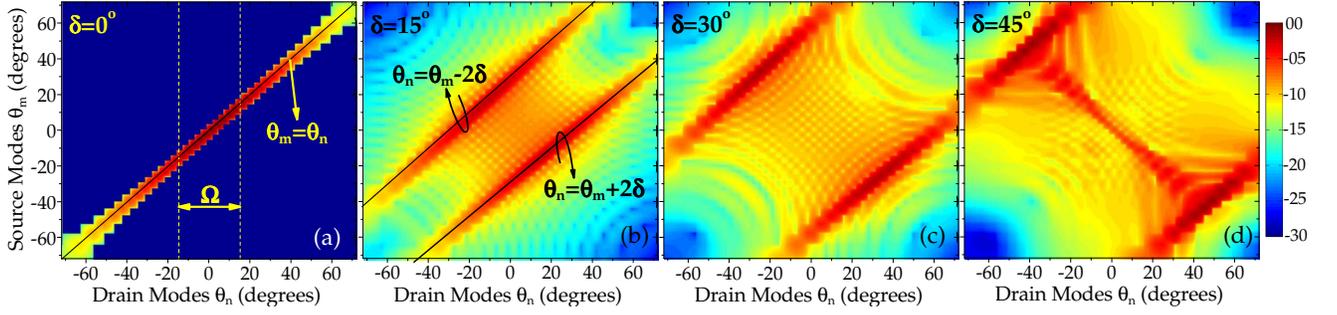}}
\caption{\footnotesize NEGF derived mode-to-mode transmission probability function $log_{2}\left[T_{negf}\left(\theta_{m},\theta_{n}\right)\right]$ for a symmetric pn junction device biased at $\epsilon_{f}=0.4eV$, for tilt angles of $\delta=0^{o},15^{o},30^{o},45^{o}$ respectively. The device width is $100nm$ and the pn transition length $D=10nm$. }
\label{tmn}
\end{figure*}
Experimentally, graphene pn junctions are created through electrical means via a top/bottom gating scheme \cite{williams07,chen08,huard07}. Carrier transport across a conventional graphene pn junction exhibits highly angular selective behavior \cite{huard07,cheianov06,katsnelson06,low08b}. For example, in a symmetric pn junction, the transmission probability in the absence of magnetic field is given by $T_{0}(k_{m})\approx e^{-\pi k_{m}^{2}D/2k_{f}}$ \cite{cheianov06}, where $k_{f/m}$ is the Fermi and transverse wave-vector respectively. When $k_{m}=0$, the transport across the pn junction would be reflectionless, an hallmark of Klein tunneling \cite{katsnelson06}. Therefore, by geometrically tilting the graphene pn junction at an angle $\delta$ as shown in Fig. \ref{figure1}, one expects that the maximum transmission now occurs for the transverse mode $k_{m}\approx k_{f}sin\delta$. Analogous to this physical situation is the problem of transport across a conventional pn junction in the presence of a magnetic field, $B$. In the latter case, one uses the Lorentz force to modify the carrier's trajectory. In the limit of large device width, one can impose the usual periodic boundary conditions and express the eigenstates as $\Psi_{m}(\bold{r})=e^{ik_{m}y}\varphi(x,B,k_{m})$ \cite{ando98}. The WKB transmission probability, $T_{B}(k_{m})$, in the presence of a B field is derived by Shytov and co-workers \cite{shytov07}. Fig. \ref{figure1}b-e plots $T_{B}(k_{m})$ for  different B field strength. Reflectionless transmission now occurs for the mode $k_{m}\approx k_{f}sin\theta_{B}$, where $\theta_{B}=sin^{-1}(v_{f}B/E)$. The polar plots exhibits the characteristic leaf-shaped feature which rotates in the presence of magnetic field. The thickness of the leaf defines the angular bandwidth ($\Omega$) of the pn junction. Decreasing $\Omega$ with increasing magnetic field is responsible for the degradation in conductance observed recently in experiments \cite{stander09}. Suppose the device is large and the effects from the boundaries are negligible, the transmission probability through a tilted junction could be described by a simple coordinate transformation, i.e. $T_{0}(\theta_{m}-\delta)$, as depicted in Fig. \ref{figure1}b-e. Both cases exhibit the signatures of a transverse current, i.e. Hall current in the magnetic field case. In this paper, we want to address the question, `Could one engineer a pseudo-Hall effect in a graphene waveguide via a tilted pn junction as shown in Fig. \ref{figure1}a?'.

\section{\label{sec:level2}Theory and Methods}

The theoretical model we employ in this work is based on the Landauer-B$\ddot{u}$ttiker formalism where the transmission function is computed within the framework of the non-equilibrium green function method \cite{datta97,haug96}.  The device Hamiltonian is described within the tight binding formalism \cite{wallace47,saito98},
\begin{eqnarray}
H=\Sigma_{i}v_{i}a_{i}^{\dagger}a_{i}+\Sigma_{ij}ta_{i}^{\dagger}a_{j} 
\label{hamil}
\end{eqnarray}
where $a_{i}^{\dagger}/a_{i}$ are the creation/destruction operator at each atomic site $i$. $v_{i}$ is the on-site potential energy, to be controlled by the top/bottom gates. Additional contributions due to local magnetization at the ribbon edges \cite{wimmer08} are not considered in this work, since our ribbon's width are relatively large. The open boundary condition for the quantum transport problem is embodied by contacts' self-energies, $\Sigma_{s/d}$, solved using an iterative scheme outlined in \cite{sancho84}. The device Green function is then computed through,
\begin{eqnarray}
G(\epsilon_{f})=\left(\epsilon_{f}-H-\Sigma_{s}-\Sigma_{d}\right)^{-1}
\label{greennn}
\end{eqnarray}
where $\epsilon_{f}$ is the Fermi energy. However, direct matrix inversion of Eq. \ref{greennn} usually proves to be computationally prohibitive. Therefore, one commonly resorts to recursive type techniques such as the recursive green function \cite{nonoyama98,anantram08} or the renormalization method \cite{grosso89}. In this work, we obtain the charge and current density of our device by combining familiar concepts from the recursive green function and the renormalization method. The detailed methodology is outlined in Appendix A. 

After solving for $G(\epsilon_{f})$, we can compute the mode resolved transmission probability function $T_{negf}^{mn}$ at $\epsilon_{f}$ via,
\begin{eqnarray}
T_{negf}^{mn}=Tr\left[\Gamma_{s,m}G\Gamma_{d,n}G^{\dagger}\right]
\label{transmis}
\end{eqnarray}
where $m/n$ denotes the modes in the source/drain contacts respectively. $\Gamma_{s/d}$ are known as the contacts' broadening functions which can be obtained from $\Sigma_{s/d}$ for each respective mode in the contacts i.e. $\Gamma_{s/d}=i2Im(\Sigma_{s/d})$. The mode-to-mode transmission function, $T^{mn}$, is a useful quantity for analyzing the transport effects in the modal space in the presence of device non-homogeneities (see e.g. \cite{low08}). Appendix B describes the procedure in obtaining the mode resolved contact self-energy $\Sigma_{s,m}$ in armchair edge graphene ribbon.

\section{\label{sec:level3}Results}

The graphene device that we investigate in this work is a 100nm wide ribbon with armchair edges. For an armchair edge ribbon, the scattering states for an incoming source mode can be written as \cite{brey06},
\begin{eqnarray}
\Psi_{m}\left(\bold{r}\right)=\frac{\bold{s}(\bold{k})}{\sqrt{2W}}\times\left\{\phi_{\bold{K}}(k_{m})-\phi_{\bold{K'}}(-k_{m})\right\}e^{ik_{x}x}
\label{cond}
\end{eqnarray}
where $\phi_{\bold{K}}(k_{m})$=$e^{i\bold{K}\cdot \bold{r}}e^{ik_{m}y}$ and $\bold{K}/\bold{K'}$ denotes the two inequivalent Dirac points in the graphene's Brillouin zone. $\bold{s}(\bold{k})$ being the pseudo-spin, describing the A/B sublattice wavefunction. Since a graphene pn junction is analogous to a negative refractive material in optics \cite{cheianov07c}, it is useful to define an angular representation for the contact modes in order to facilitate discussion using simple ray analysis. For the source/drain modes, we define $\theta_{m/n}=sin^{-1}(k_{m/n}/k_{f})$ respectively, where $n$ labels the modes in the drain.  Fig. \ref{tmn} plots the mode-to-mode transmission function $T_{negf}^{mn}$ for various tilted pn junctions ($\delta=0^{o},15^{o},30^{o},45^{o}$) biased symmetrically with a built-in potential of $0.8eV$ i.e. $\epsilon_{f}=0.4eV$ on each side. When the pn interface tilt angle is zero, the solutions satisfy the `Snell law' given by $\theta_{m}=\theta_{n}$ (see Fig. \ref{tmn}a). When $\delta\neq 0$, the solutions are generally described by:
\begin{eqnarray}
\theta_{n}=\left\{
\begin{array}{ccc}
\theta_{m}+2\delta & : & \bold{K} \\
\theta_{m}-2\delta & : & \bold{K'} 
\end{array}
\right.
\label{thetasdeq}
\end{eqnarray}
as demonstrated in Fig. \ref{tmn}b. Each source mode is an equal weight superposition of scattering states from $\bold{K}$ and $\bold{K'}$ valleys propagating in $\pm\theta_{m}$ direction respectively. When $\delta\neq 0$, these two scattering states get scattered differently, ending up in two different drain modes according to Eq. \ref{thetasdeq}. However, when $\delta$ exceeds a maximum tilt, i.e. $\delta_{max}$, the solutions described by Eq. \ref{thetasdeq} fall outside of the available drain modes and new solutions emerge (see Fig. \ref{tmn}d). We find,
\begin{eqnarray}
\delta_{max} \approx \tfrac{1}{2}max\left(\theta_{n}\right)
\label{equality1}
\end{eqnarray}
which can be easily deduced from Fig. \ref{tmn} (for a 100nm armchair edge ribbon, max$\left(\theta_{n}\right)\approx 72^{o}$). We will revisit this point later. 

\begin{figure}[t]
\centering
\scalebox{0.47}[0.47]{\includegraphics*[viewport=130 275 520 480]{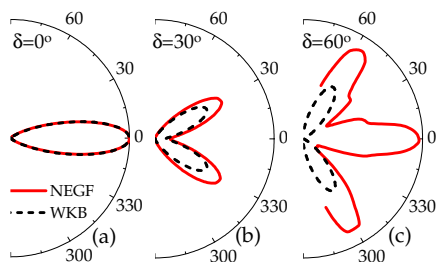}}
\caption{\footnotesize (a-c) Polar plots of the transmission probability $T_{negf}(\theta_{m})=\Sigma_{n}T_{negf}^{mn}$ for different $\delta$, where $T_{negf}^{mn}$ is computed using the same device parameters as for Fig. \ref{tmn}. The WKB plots are obtained using the usual WKB formula \cite{cheianov06} (see text)  }
\label{F3}
\end{figure}

Summing over the drain modes, we obtained the transmission probability due to an incoming mode from the source: $T_{negf}(\theta_{m})=\Sigma_{n}T_{negf}^{mn}$. Fig. \ref{F3} is the polar plot of $T_{negf}(\theta_{m})$ for symmetric pn junction devices with different $\delta$. The WKB plots are obtained using the usual WKB formula \cite{cheianov06}, and treating the two scattering states from each mode independently. When $\delta$ increases gradually from zero, the single leaf evolves into a doublet leaf structure as shown in Fig. \ref{F3}b. This is because Klein tunneling which originally occurs for the mode $\theta_{m}\approx 0$, now occurs for the modes $\theta_{m}\approx \pm\delta $. For the latter, one would expect the maximum transmission probability to be $\approx$ $\tfrac{1}{2}$, since only one of the pair of scattering states from each mode satisfy the condition for Klein tunneling.  However, the NEGF result deviates from this simple picture, showing a notably higher maximum transmission than $\tfrac{1}{2}$. The reason for this discrepency is due to multiple scattering with the sidewall i.e. sidewall enhanced transmission (SWET). With further increase in $\delta$, this doublet leaf structure evolves into some triplet leaf feature (new solutions arise when  $\delta>\delta_{max}$) and eventually the polar plot becomes noisy (not shown). Increasing of $\delta$ extends the physical longitudinal distance of the tilted gate, thereby enhancing the mixing of the various transverse modes.  

\begin{figure}[t]
\centering
\scalebox{0.43}[0.43]{\includegraphics*[viewport=130 175 690 490]{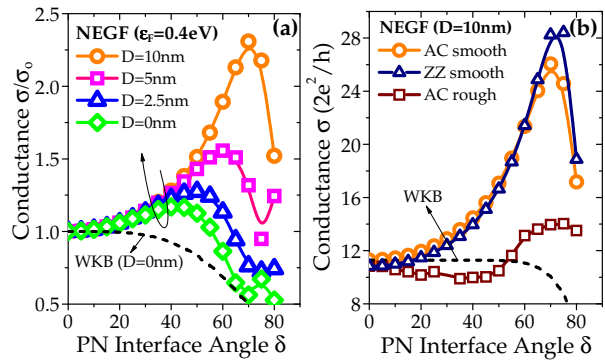}}
\caption{\footnotesize (a) Normalized conductance, $\sigma/\sigma_{0}$, of a pn junction as a function of the tilt angle $\delta$, computed for various values of transition length $D$ and biased symmetrically at $\epsilon_{f}=0.4eV$. $\sigma_{0}$ is defined as the conductance when $\delta=0$, which by construction means $\sigma/\sigma_{0}=1$ when $\delta=0$. (b) Absolute conductance for the $D=10nm$ device with (i) perfectly smooth armchair edges (round symbol), roughened armchair edges i.e. RMS roughness of 1 atomic layer (square symbol) and perfectly smooth zigzag edges (triangle symbol). Dashed line is the result from simple WKB model.}
\label{cond}
\end{figure}

\subsection{\label{seccond}Junction conductance}

Fig. \ref{cond}a plots the pn junction normalized conductance ($\sigma/\sigma_{o}$) as a function of tilt angle $\delta$ for different values of $D$, where $\sigma_{o}$ is the conductance when $\delta=0^{o}$. The following key observations can be made; (i) $\sigma/\sigma_{0}$ exhibits an initial increase with $\delta$ and then decreases prominently when $\delta$ exceeds a threshold angle, herein denoted as $\delta_{th}$, and (ii) the occurence of $\delta_{th}$ can be delayed by employing a larger $D$. We also checked that the same trends hold true for devices with different widths, $\epsilon_{f}$ and edge configurations (i.e. zigzag edge ribbon shown in Fig. \ref{cond}b). The initial increase in conductance with $\delta$ is a SWET phenomenon whose signature becomes more prominent with larger $\delta$.  We shall discuss the plausible explanation for the existance of $\delta_{th}$, beyond which $\sigma/\sigma_{0}$ degrades. In the transmission polar plot (Fig. \ref{figure1}), increasing $\delta$ rotates the leaf by a similar amount. The threshold of conductance degradation occurs at large enough $\delta$ such that some of the states within the angular bandwidth would be back reflected into the source i.e.
\begin{eqnarray}
\delta_{th}\approx \tfrac{1}{2}\left(\pi-\Omega\right)
\label{equality1}
\end{eqnarray}
where $\Omega$ is the angular bandwidth of the transmission function. Since $\Omega$ is larger for smaller $D$, $\delta_{th}$ is also smaller. This explains the general trend we observe in our numerical calculation in Fig. \ref{cond}. Let us compute $\delta_{th}$ for the set of results in Fig. \ref{cond}a. Defining the $\Omega$ to be the bandwidth where transmission probability is $>0.5$, we have $\Omega\approx 90^{o},75^{o},49^{o},31^{o}$ when $D=0,2.5,5,10nm$ respectively. This yields $\delta_{th}\approx 45^{o},53^{o},66^{o},74^{o}$ respectively, in good agreement with the numerical result we obtained in Fig. \ref{cond}a. 

\begin{figure}[t]
\centering
\scalebox{0.63}[0.63]{\includegraphics*[viewport=190 90 565 566]{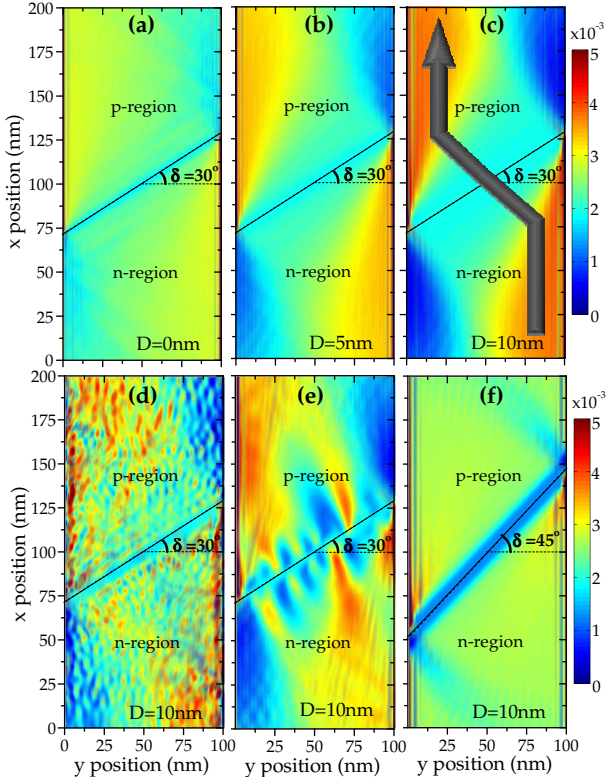}}
\caption{\footnotesize (a-c) shows the non-equilibrium real space longitudinal current density for pn devices with $D=0nm,5nm,10nm$ respectively. All devices considered here have $\delta=30^{o}$ and $\epsilon_{f}=0.4eV$. (d-e) are the same device as (c) excepts with sidewall roughness and pn interface roughness respectively. The RMS roughness are 1 atomic layer and 1.7nm respectively. (f) is the device with $D=10nm$ and $\delta=45^{o}$ with no disorder. }
\label{current}
\end{figure}
For device with $D=10nm$, the junction conductance at $\delta=70^{o}$ can exceed twice its value at $\delta=0^{o}$, as shown in Fig. \ref{cond}a. From a device perspective, this means that one could deliberately tilt the interface angle to enhance the on-state current. This could be useful for engineering a band-to-band tunneling transistor \cite{appenzeller04}. Next, we examine the robustness of this effect in the presence of sidewall disorder. Fig. \ref{cond}b plots the absolute junction conductance of the $D=10nm$ device for the case with a perfect sidewall and one where the sidewall exhibits a RMS roughness of one atomic layer. Evidently, the SWET phenomenon is highly sensitive to the characteristic of the sidewall. Therefore, chemically derived graphene ribbons \cite{li08} with smooth edges are needed to experimentally observe these large conductance modulation with tilt angle. Device widths of the same order as the carrier's phase coherence length $L_{\phi}$ is also required for SWET to occur. From Fabry Perot experiments \cite{young08}, $L_{\phi}\approx 100nm$ is expected.

\subsection{\label{spatialc}Spatial current distribution}

Fig. \ref{current}a-f shows the non-equilibrium spatial current intensity plots of a tilted graphene pn junction for different transition lengths and tilt angle. The current profile populates preferentially along the sides of the device, analogous to the Hall effect.  The following key observations can be made about Fig. \ref{current}a-f; (i) edge populated current is observed to be more prominent with increasing $D$ and (ii) this effect can be enhanced by increasing $\delta$ until $\delta$ exceeds a certain angle (see Fig. \ref{current}f). The former effect is attributed to the suppression of normal modes current (i.e. $T_{negf}(\theta_{m}\approx 0)$) as a result of larger $D$. 

\begin{figure}[t]
\centering
\scalebox{0.5}[0.5]{\includegraphics*[viewport=120 205 700 360]{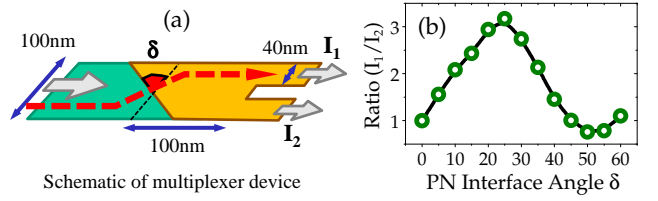}}
\caption{\footnotesize (a) Schematic of a multiplexer device where the drain contact is partitioned into two via graphene stubs. (b) Ratio of the current through the two drain terminals as a function of $\delta$. Parameters assumed are $D=10nm$, $\epsilon_{f} = 0.4eV$ and under small source drain bias.}
\label{multiplexer}
\end{figure}

The appearance of edge populated current is a direct consequence of the negative refractive index property of the graphene pn junction. Each propagating mode from the source follows a classical trajectory into the drain contact according to Eq. \ref{thetasdeq}. For example, when $\delta=15^{o}$, the incoming source modes $\theta_{m}\approx\pm 15^{o}$ would contribute the most current (see Fig. \ref{F3}b and Fig. \ref{tmn}b). The mode $\theta_{m}\approx 15^{o}$ would end up in the drain modes $\theta_{n}\approx 45^{o}(\bold{K})$ and $\theta_{n}\approx -15^{o}(\bold{K}')$, where the latter scattering state has a larger current contribution. $(...)$ indicates the valley in which the incoming scattering state is residing. On the other hand, $\theta_{m}\approx -15^{o}$ would end up in the drain modes $\theta_{n}\approx 15^{o}(\bold{K})$ and $\theta_{n}\approx -45^{o}(\bold{K}')$, where the former has a larger contribution. Both scattering states, $\theta_{n}\approx -15^{o}(\bold{K}')$ and $\theta_{n}\approx 15^{o}(\bold{K})$, are propagating in the same direction, since $\bold{K}=-\bold{K}'$. This leads to the effect of pseudo-Hall current. However, when $\delta>\delta_{max}$, new scattering states which do not follow the classical trajectories described by the `Snell law' (i.e. Eq. \ref{thetasdeq}) arises. They are responsible for overwhelming the edge populated currents, resulting in their disappearance for the device with $\delta=45^{o}$ (see Fig. \ref{current}f). Unlike the SWET, the phenomenon of edge populated current is fairly robust against various disorder such as edge roughness and pn interface roughness as shown in Fig. \ref{current}d-e. One should be able to measure this effect experimentally \footnote{One should be careful with the placement of metal contacts, which would modify the potential landscape of the underlying graphene and results in reflections at the metal/graphene interface \cite{miao07}. Non-invasive metallic probes through graphene stubs would probably be a prefered way to contact the device \cite{huard08r}. }. 

We consider a possible experimental setup that allows direct access to the proposed effect as shown in Fig. \ref{multiplexer}a. The drain is partitioned into two contacts through an upper/lower stub, with currents denoted as $I_{1}$/$I_{2}$ respectively. Fig. \ref{multiplexer}b plots the ratio $I_{1}$/$I_{2}$ as a function of $\delta$. The current asymmetry has an optimum value of $300\%$ at $\delta\approx 25^{o}$ as shown. Through further device optimization, one should be able to engineer a device with a larger asymmetry ratio and achieve a semi-unipolar behavior through $I_{2}$. 

In conclusion, we had peformed a numerical study of a tilted graphene pn junction, provides a detailed physical understanding of its transport properties, and highlighted the possibility of manipulating the current to flow along the edges of the waveguide. \\

$\bold{Acknowledgement}$ TL gratefully acknowledges the financial support from Nanoelectronics Research Initiative and the computational resources provided by Network for Computational Nanoelectronics. We gratefully acknowledge the useful discussions with Mark Lundstrom, Supriyo Datta and Dmitri Nikonov.

\appendix
\section{\label{sec:appen} Renormalization and recursive methods}

\begin{figure}[t]
\centering
\scalebox{0.4}[0.4]{\includegraphics*[viewport=80 160 720 490]{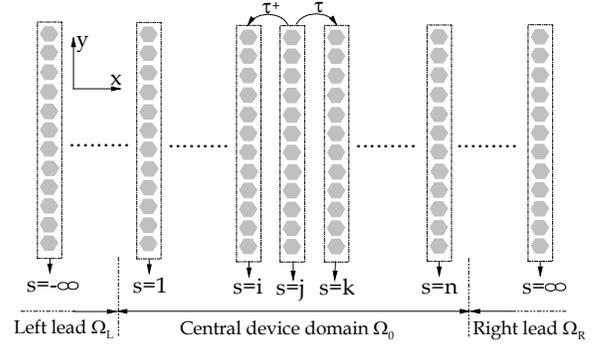}}
\caption{\footnotesize Schematic on a graphene ribbon with armchair edges. Each slice of supercell consist of an intra-cell interation denoted by $\alpha$ and a right/left neighboring cell interaction represented by $\tau/\tau^{\dagger}$ respectively.   }
\label{negf1}
\end{figure}

This appendix documents the procedure we used for the computation of spatial charge/current density profiles in the device. We consider a graphene ribbon with armchair edges as illustrated in Fig. \ref{negf1}. Device is infinite along x, the transport direction, and each supercell is represented by the dotted rectangular box. The Hamiltonian, $H$, describing the graphene ribbon is formulated by treating only the nearest-neighbor interaction between the pz-orbitals \cite{wallace47,saito98}. Usually, this coupling energy is assumed to be $t_{c}=3eV$. By the same token, the supercell would only interact with the adjacent supercells. The interation of a supercell with its neighboring cell on the right/left is represented by $\tau/\tau^{\dagger}$ respectively, while the intra-cell interation is denoted by $\alpha$. $\tau$ and $\alpha$ are matrices of size $n_{s}\times n_{s}$, where $n_{s}$ is the number of basis functions in a supercell. $H$ is the sum of these coupling energies and the electrostatic potential $U(\bold{r})$.

From a practical point of view, we are only interested in the scattering solutions within the central device domain, denoted by $\Omega_{0}$. Let $H_{L}$, $H_{0}$ and $H_{R}$ be the Hamiltonian description of $\Omega_{L}$, $\Omega_{0}$ and $\Omega_{R}$ respectively. The interaction between the $H_{L}$ and $H_{0}$ block is denoted by $\tilde{\tau}$. Through simple algebras, one can write the retarded green function at $\epsilon_{f}$ (Fermi energy) in $\Omega_{0}$ as follows \cite{datta97},
\begin{eqnarray}
G(\bold{r},\bold{r}')=\left((\epsilon_{f}+i\eta)I-H_{0}- \Sigma_{s}-\Sigma_{d}  \right)^{-1}\equiv A^{-1}
\label{greenfunction}
\end{eqnarray}
$\Sigma_{s/d}$ are known as the contact retarded self-energy and are defined as follows, 
\begin{eqnarray}
\begin{array}{ccc}
\Sigma_{s}=\tilde{\tau}^{\dagger}g_{L}^{r}\tilde{\tau} & & g_{L}^{r}=\left((\epsilon_{f}+i\eta)I-H_{L}\right)^{-1}\\
\Sigma_{d}=\tilde{\tau} g_{R}^{r}\tilde{\tau}^{\dagger}  & & g_{R}^{r}=\left((\epsilon_{f}+i\eta)I-H_{R}\right)^{-1}
\end{array}
\label{selfenerr}
\end{eqnarray}
The numerics for Eq. \ref{selfenerr} immediately becomes tractable when one notice that we only need the elements of $g_{L/R}^{r}$ which are adjacent to the $\Omega_{R}\cap \Omega_{0}$ and $\Omega_{L}\cap \Omega_{0}$ boundaries. These `surface elements', which are a smaller subset of $g_{L/R}^{r}$, are usually denoted by the surface green function matrices $g_{L/R}^{s}$ of size $n_{s}\times n_{s}$. An iterative scheme is commonly used to compute $g_{L/R}^{s}$ \cite{sancho84}. Once the contact retarded self-energies are determined, the device green function in Eq. \ref{greenfunction} can be computed by directly inverting the matrix $A$, if computational resource is not a limiting factor. However, it usually prove to be computationally prohibitive. Therefore, one commonly resorts to recursive type techniques such as the recursive green function \cite{nonoyama98,anantram08} or the renormalization method \cite{grosso89}. In this appendix, we outlined a methodology which combines familiar concepts from the recursive green function and the decimation method in the computation of the various non-equilibrium green functions, from which we can obtain the charge and current density of our device. 

Suppose that we are only interested in the real space resolved charge and current density for the supercell $j$ as shown in Fig. \ref{negf1}. The first step involves getting rid of the slices $s=2,3,\ldots,h,l,\ldots,n-2,n-1$ from the system of equations stipulated in Eq. \ref{greenfunction}, made possible by the trigonal nature of matrix $A$. Eq. \ref{greenfunction} now becomes $\bar{G}\bar{A}=I$, where the LHS of this equation is explicitly written as,
\begin{eqnarray}
\left[
\begin{array}{ccccc}
g_{11} & g_{1i} & g_{1j} & g_{1k} & g_{1n}\\
g_{i1} & g_{ii} & g_{ij} & g_{ik} & g_{in}\\
g_{j1} & g_{ji} & g_{jj} & g_{jk} & g_{jn} \\
g_{k1} & g_{ki} & g_{kj} & g_{kk} & g_{kn}\\
g_{n1} & g_{ni} & g_{nj} & g_{nk} & g_{nn}
\end{array}
\right]\left[
\begin{array}{ccccc}
a_{11} & a_{1i} & 0 & 0 & 0\\
a_{i1} & a_{ii} & a_{ij} & 0 & 0\\
0 & a_{ji} & a_{jj} & a_{jk} & 0 \\
0 & 0 & a_{kj} & a_{kk} & a_{kn}\\
0 & 0 & 0 & a_{nk} & a_{nn}
\end{array}
\right]
\label{newA}
\end{eqnarray}
In this work, the matrix elements of $\bar{A}$ are systematically derived. The elements not affected by the decimation process are $a_{jj}=[A]^{j}_{j}$, $a_{ij}=[A]^{i}_{j}$, $a_{ji}=[A]^{j}_{i}$, $a_{jk}=[A]^{j}_{k}$ and $a_{kj}=[A]^{k}_{j}$ (the upper/lower index denotes row/column respectively). $a_{ii}$ and $a_{i1}$ are obtained through a set of recursive formulae. We began with the initialization $a^{0}_{ii}=[A]^{i}_{i}$ and $a^{0}_{i1}=\tau^{\dagger}$. The recursive formulae for $a_{ii}$ and $a_{i1}$ are,
\begin{eqnarray}
\nonumber
a_{ii}^{u}&=&a_{ii}^{u-1}-a_{i1}^{u-1}p^{u}_{l}\left(a_{i1}^{u-1}\right)^{\dagger}\\
\nonumber
a_{i1}^{u}&=&-a_{i1}^{u-1}p^{u}_{l}\tau^{\dagger}\\
p^{u}_{l}&=&\left([A]^{j-u-1}_{j-u-1}-\tau p^{u-1}_{l}\tau^{\dagger} \right)^{-1}
\label{recur1}
\end{eqnarray}
where $p^{u}_{l}=0$. The desired solutions are $a_{ii}=a_{ii}^{j-3}$ and $a_{i1}=a_{i1}^{j-3}$. Note that a different set of recursive formula is needed if the intercell coupling is different for each supercell. $a_{kk}$ and $a_{kn}$ are obtained through a similar set of recursive formulae. We began with the initialization $a^{0}_{kk}=[A]^{k}_{k}$ and $a^{0}_{kn}=\tau$. The recursive formulae for $a_{kk}$ and $a_{kn}$ are,
\begin{eqnarray}
\nonumber
a_{kk}^{u}&=&a_{kk}^{u-1}-a_{kn}^{u-1}p^{u}_{r}\left(a_{kn}^{u-1}\right)^{\dagger}\\
\nonumber
a_{kn}^{u}&=&-a_{kn}^{u-1}p^{u}_{r}\tau\\
p^{u}_{r}&=&\left([A]^{j-u-1}_{j-u-1}-\tau^{\dagger} p^{u-1}_{r}\tau \right)^{-1}
\label{recur2}
\end{eqnarray}
where $p^{u}_{l}=0$. The desired solutions are $a_{kk}=a_{kk}^{n-j-2}$ and $a_{kn}=a_{kn}^{n-j-2}$. $a_{11}$ and $a_{1i}$ are obtained through a similar set of recursive formulae. We began with the initialization $a^{0}_{11}=[A]^{1}_{1}$ and $a^{0}_{1i}=\tau$. The recursive formulae are;
\begin{eqnarray}
\nonumber
a_{11}^{u}&=&a_{11}^{u-1}-a_{1i}^{u-1}q^{u}_{l}\left(a_{1i}^{u-1}\right)^{\dagger}\\
\nonumber
a_{1i}^{u}&=&-a_{1i}^{u-1}q^{u}_{l}\tau\\
q^{u}_{l}&=&\left([A]^{1+u}_{1+u}-\tau^{\dagger} q^{u-1}_{l}\tau \right)^{-1}
\label{recur3}
\end{eqnarray}
where $q^{u}_{l}=0$. The desired solutions are $a_{11}=a_{11}^{j-3}$ and $a_{1i}=a_{1i}^{j-3}$. $a_{nn}$ and $a_{nk}$ are obtained through a similar set of recursive formulae. We began with the initialization $a^{0}_{nn}=[A]^{n}_{n}$ and $a^{0}_{nk}=\tau^{\dagger}$. The recursive formulae are;
\begin{eqnarray}
\nonumber
a_{nn}^{u}&=&a_{nn}^{u-1}-a_{nk}^{u-1}q^{u}_{r}\left(a_{nk}^{u-1}\right)^{\dagger}\\
\nonumber
a_{nk}^{u}&=&-a_{nk}^{u-1}q^{u}_{r}\tau^{\dagger}\\
q^{u}_{r}&=&\left([A]^{n-u}_{n-u}-\tau q^{u-1}_{r}\tau^{\dagger} \right)^{-1}
\label{recur4}
\end{eqnarray}
where $q^{u}_{r}=0$. The desired solutions are $a_{nn}=a_{nn}^{n-j-2}$ and $a_{nk}=a_{nk}^{n-j-2}$. Performing the recursive procedure in Eq. \ref{recur1}-\ref{recur4}, one can then obtain the full information of the matrix $\bar{A}$. 

We are now ready to compute the charge and current density for the supercell $j$. The key quantity is the electron correlation function given by,
\begin{eqnarray}
\nonumber
G^{n}&=&G\left(\Sigma^{in}\right)G^{\dagger}
\label{negfeq}
\end{eqnarray}
where $\Sigma^{in}=\Sigma^{in}_{s}+\Sigma^{in}_{d}$ are known as the in-scattering self energies. It is given by,
\begin{eqnarray}
\Sigma^{in}_{s/d}&=&if_{s/d}\left(\Sigma_{s/d}-\Sigma_{s/d}^{\dagger}\right)
\label{dyson}
\end{eqnarray}
where $f_{s/d}$ are the Fermi occupation factor in the source and drain contacts. 

The key concepts in recursive solution of $G$ can now be employed to solve Eq. \ref{newA}. Specifically, we only require the solution to $g_{jj}$, $g_{jn}$, $g_{kn}$, $g_{j1}$ and $g_{k1}$ for reasons that would be apparent later. We would need the following recursive formulae,
\begin{eqnarray}
\nonumber
[G]^{q}_{q}&=&\Omega^{q}-\Omega^{q}[A]^{q}_{q+1}[G]^{q+1}_{q}\\
\nonumber
\left[G\right]^{q}_{q+r}&=&-\Omega^{q}[A]^{q}_{q+1}[G]^{q+1}_{q+r}\\
\Omega^{v+1}&=&\left([\bar{A}]^{v+1}_{v+1}-[\bar{A}]^{v+1}_{v}\Omega^{v}[\bar{A}]^{v}_{v+1}\right)^{-1}
\label{recurgreenfor2}
\end{eqnarray}
where $\Omega^{0}=0$ and it yields us $\Omega^{5}=g_{nn}$. We can then arrive at the following results,
\begin{eqnarray}
\nonumber
g_{kn}&=&-\Omega^{4}[\bar{A}]^{4}_{5}g_{nn}\\
\nonumber
g_{jn}&=&-\Omega^{3}[\bar{A}]^{3}_{4}g_{kn}\\
\nonumber
g_{kk}&=&\Omega^{4}-\Omega^{4}[\bar{A}]^{4}_{5}g_{kn}^{T}\\
\nonumber
g_{jj}&=&\Omega^{3}-\Omega^{3}[\bar{A}]^{3}_{4}\left(\Omega^{3}[\bar{A}]^{3}_{4}g_{kk}\right)^{T}\\
\nonumber
g_{j1}&=&(-1)^{2}\left(\Omega^{1}[\bar{A}]^{1}_{2}\Omega^{2}[\bar{A}]^{2}_{3}g_{jj}\right)^{T}\\
g_{k1}&=&(-1)^{3}\left(\Omega^{1}[\bar{A}]^{1}_{2}\Omega^{2}[\bar{A}]^{2}_{3}\Omega^{3}[\bar{A}]^{3}_{4}g_{kk}\right)^{T}
\label{formulag}
\end{eqnarray}
With these block elements information of $G$, we are now ready to compute the charge and current density.

We are interested in the electron density, $n(\bold{r})$, of the supercell $j$ given by,
\begin{eqnarray}
\nonumber
\left[G^{n}\right]^{j}_{j}&=&\left[G\right]^{j}_{1}\left[\Sigma^{in}_{s}\right]^{1}_{1}\left[G^{\dagger}\right]^{1}_{j}+\left[G\right]^{j}_{n}\left[\Sigma^{in}_{d}\right]^{n}_{n}\left[G^{\dagger}\right]^{n}_{j}\\
&=&g_{j1}\left[\Sigma^{in}_{s}\right]^{1}_{1}g_{j1}^{\dagger}+g_{jn}\left[\Sigma^{in}_{d}\right]^{n}_{n}g_{jn}^{\dagger}
\label{correlj}
\end{eqnarray}
where we had make use of the fact that $\Sigma^{in}_{s/d}$ are non-zero only for $j=1,n$ slices respectively. The current density, $j(\bold{r})$, flowing between the supercell $j$ and $j+1$ is computed via,
\begin{eqnarray}
j(\bold{r})=\frac{2q}{h}\left(\left[A\right]^{j}_{j+1}\left[G^{n}\right]^{j+1}_{j}-\left[A\right]^{j+1}_{j}\left[G^{n}\right]^{j}_{j+1}\right)
\label{current1}
\end{eqnarray}
where,
\begin{eqnarray}
\nonumber
\left[G^{n}\right]^{j+1}_{j}&=&\left[G\right]^{j+1}_{1}\left[\Sigma^{in}_{s}\right]^{1}_{1}\left[G^{\dagger}\right]^{1}_{j}+\left[G\right]^{j+1}_{n}\left[\Sigma^{in}_{d}\right]^{n}_{n}\left[G^{\dagger}\right]^{n}_{j}\\
&=&g_{k1}\left[\Sigma^{in}_{s}\right]^{1}_{1}g_{j1}^{\dagger}+g_{kn}\left[\Sigma^{in}_{d}\right]^{n}_{n}g_{jn}^{\dagger}\\
\nonumber
\left[G^{n}\right]^{j}_{j+1}&=&\left[G\right]^{j}_{1}\left[\Sigma^{in}_{s}\right]^{1}_{1}\left[G^{\dagger}\right]^{1}_{j+1}+\left[G\right]^{j}_{n}\left[\Sigma^{in}_{d}\right]^{n}_{n}\left[G^{\dagger}\right]^{n}_{j+1}\\
&=&g_{j1}\left[\Sigma^{in}_{s}\right]^{1}_{1}g_{k1}^{\dagger}+g_{jn}\left[\Sigma^{in}_{d}\right]^{n}_{n}g_{kn}^{\dagger}
\label{correlj2}
\end{eqnarray}
Therefore, we have completed our procedure in computing the charge and current density for the supercell $j$. At any time, our numerical procedure only requires direct matrix inversion of size $n_{s}\times n_{s}$, where $n_{s}$ is the number of basis functions in a supercell. By parallelizing the computations, we can compute the charge and current density for any number of supercells within the device domain. In our work, we had computed for a dozens of supercell to give us the required spatial resolution of the charge and current density related graphical plots in the main paper.

\section{\label{sec:appen} Mode resolved contact self energy}

The mode-to-mode transmission function, $T^{mn}$, is a useful quantity for analyzing the transport effects in the modal space in the presence of device non-homogeneities (see e.g. \cite{low08}). It is given by $T^{mn}=Tr\left[\Gamma_{s,m}G\Gamma_{d,n}G^{\dagger}\right]$, where $m/n$ denotes the modes in the source/drain contacts respectively. $\Gamma_{s/d}$ are known as the contacts' broadening functions which can be obtained from $\Sigma_{s/d}$ for each respective modes in the contacts i.e. $\Gamma_{s/d}=i2Im(\Sigma_{s/d})$. This appendix describe the procedure in obtaining the mode resolved contact self-energy $\Sigma_{s,m}$ in armchair edge graphene ribbon.

In armchair ribbon, the analytical solutions of the wavefunction and energy dispersion is known analytically \cite{zheng07}. One could construct a unitary operator $V$ which perform the transformation from real space to mode space. Zhao and Guo \cite{zhao09} outlined the recipe for doing so. For mode $m$, its propagation along the lattice chain could be described by an on-site and coupling matrix $\alpha$ and $\beta$ respectively,
\begin{eqnarray}
\begin{array}{cc}
\alpha=\left[
\begin{array}{cccc}
0 & \omega_{m} & 0 & 0 \\
\omega_{m} & 0 & t_{c} & 0 \\
0 & t_{c} & 0 & \omega_{m}\\
0 & 0 & \omega_{m} & 0 \\
\end{array}
\right] & 
\beta=\left[
\begin{array}{cccc}
0 & 0 & 0 & 0 \\
0 & 0 & 0 & 0 \\
0 & 0 & 0 & 0 \\
t_{c} & 0 & 0 & 0 \\
\end{array}
\right]
\end{array}
\end{eqnarray}
where $\omega_{m}=2t_{c}cos\left(m\pi/(2L+1)\right)$, $L$ being the number of carbon layers along the width direction. In the paper, the angular representation for mode $m$ is given by $\theta_{m}=sin^{-1}\left[(t_{c}-\omega_{m})/\epsilon_{f}\right]$. The self-energy for the semi-infinite leads of this lattice chain is denoted by $\Xi_{m}$ and could be computed rather inexpensively via the usual technique described in \cite{sancho84} or analytically as discussed in \cite{zhao09}. Finally, the real space form of the mode-resolved self-energy is given by,
\begin{eqnarray}
\Sigma_{s,m}=V\left(\Xi_{m}\otimes I_{m}\right)V^{\dagger}
\end{eqnarray}
$V$ is a $n_{s}\times n_{s}$ unitary matrix, whose elements values are assigned as described in \cite{zhao09}. $I_{m}$ is a $n_{s}/4\times n_{s}/4$ matrix with elements given by $I_{m}(i,j)=\delta_{i,m}\delta_{j,m}$. $\Sigma_{s,m}$ is therefore of size $n_{s}\times n_{s}$. To ensure that the procedure is correct, we check the following sum rule,
\begin{eqnarray}
\Sigma_{s}=\Sigma_{s,1}+\Sigma_{s,2}+\Sigma_{s,3}+\ldots
\end{eqnarray}
This completes the objective of this appendix.

\end{document}